\documentclass[letterpaper, 10 pt, conference]{ieeeconf}  

\IEEEoverridecommandlockouts                              

\usepackage{xcolor, soul}
\usepackage{graphics}
\usepackage{amsmath}
\usepackage{graphicx}
\usepackage{comment}
\usepackage{cite}
\usepackage[autostyle, english = american]{csquotes}
\MakeOuterQuote{"}
\IEEEoverridecommandlockouts 

\title{\LARGE \bf
Appropriateness of LLM-equipped Robotic Well-being Coach Language in the Workplace: A Qualitative Evaluation}

\author{Micol Spitale$^{1, 2}$, Minja Axelsson$^{1}$ and Hatice Gunes$^{1}$
\thanks{$^{1}$ Department of Computer Science and Technology, University of Cambridge, UK, $^{2}$ Department of Electronic, Information, and Bio-engineering, Politecnico di Milano, Italy.  
        {\tt\small micol.spitale@polimi.it}}%
}

\begin{document}

\maketitle
\thispagestyle{empty}
\pagestyle{empty}

\begin{abstract}


Robotic coaches have been recently investigated to promote mental well-being in various contexts such as workplaces and homes. With the widespread use of Large Language Models (LLMs), HRI researchers are called to consider language appropriateness when using such generated language for robotic mental well-being coaches in the real world. Therefore, this paper presents the first work that investigated the language appropriateness of robot mental well-being coach in the workplace. 
To this end, we conducted an empirical study that involved 17 employees who interacted over 4 weeks with a robotic mental well-being coach equipped with LLM-based capabilities. After the study, we individually interviewed them and we conducted a focus group of 1.5 hours with 11 of them. 
The focus group consisted of: i) an ice-breaking activity, ii) evaluation of robotic coach language appropriateness in various scenarios, and iii) listing \textit{should\emph{s}} and\textit{ shouldn't\emph{s}} for designing appropriate robotic coach language for mental well-being. 
From our qualitative evaluation, we found that a language-appropriate robotic coach should (1) ask deep questions which explore feelings of the coachees, rather than superficial questions, (2) express and show emotional and empathic understanding of the context, and (3) not make any assumptions without clarifying with follow-up questions to avoid bias and stereotyping. 
These results can inform the design of language-appropriate robotic coach to promote mental well-being in real-world contexts.
\end{abstract}

\section{INTRODUCTION}

The last year has been characterised by a ground-breaking advancement in Large Language Models (LLMs) that has revolutionised several research fields \cite{thirunavukarasu2023large, amin2023will}, including Human-Robot Interaction (HRI) research \cite{zhang2023large}. Robots have been increasingly introduced in sensitive application contexts -- such as mental health \cite{spitale2023robotic, abbasi2022can}, therapy \cite{spitale2023using, scassellati2012robots}, and elderly care \cite{carros2020exploring}. In such sensitive contexts, designing appropriate robot language is extremely important. 
As HRI researchers, we are called to consider the appropriateness of using LLM-generated language in robots deployed in the real world, and the social and ethical implications of their language when interacting with people.

Within the HRI literature, \textit{robotic well-being coaches} have been recently investigated with the aim of promoting mental well-being in homes \cite{jeong2023robotic}, university accommodations\cite{jeong2023deploying}, and workplaces \cite{spitale2023robotic, spitale2023longitudinal}. However, the appropriateness of these robotic coaches' language, generated using LLMs, have not been analyzed. 
In the past, such appropriateness analyses have been undertaken in other HRI applications such as robotic movement planning \cite{liu2023llm}. Past studies have shown how to design an appropriate prompt to guide researchers in generating high-level motion planning \cite{song2023llm, huang2022language, cui2023no}.
For example, Huang et al. \cite{huang2022language} have used LLMs to 
explore the feasibility of decomposing natural language expressions for high-level tasks (such as "make breakfast") into a designated set of 
 executable steps (for instance, "open fridge"); while Song et al. \cite{song2023llm} proposed an effective way to use LLMs as planners to execute complex tasks based on natural language instructions within a visually perceived environment. 
Our work aims to extend such analysis of appropriateness to the language robots use during HRI---specifically to the design of language-appropriate \textit{robotic mental well-being coaches} in real-world settings. 

This paper presents the first study that examines the \textbf{appropriateness of robotic mental well-being coach language} in the workplace. 
To this end, we conducted an empirical study with 17 employees who interacted with an LLM-equipped robotic mental well-being coach over 4 weeks in their workplace \cite{spitale2024NEW}. After the study, we first interviewed them individually, and then we organised a focus group that involved 11 of them. 
This focus group included: 1) an ice-breaking activity, where participants were asked to describe the robotic coach with five adjectives; 2) language appropriateness evaluation, where participants were asked to assess the appropriateness of robotic coach language in seven different scenarios; and 3) listing \textit{should\emph{s}} and \textit{shouldn't\emph{s}}, where participants were asked to list what a robotic coach should and should not say and how it should behave when delivering well-being practices.

Our results show that a language-appropriate robotic well-being coach should (1) ask deep questions which explore feelings of the coachees, (2) express and show emotional and empathic understanding of the context, and (3) not make any assumptions without clarifying with follow-up questions to avoid bias and behaviours that enforce stereotypes.
Our analysis and results contribute towards understanding how to design language-appropriate robotic coaches to promote mental well-being in adults in their workplace.

\section{RELATED WORK}
\label{sec:rw}

\subsection{Appropriate LLM-equipped Agent Language}

Designing embodied agent (e.g., robot, virtual agent) language that is appropriate for a specific context can be a challenging task \cite{wang2023voyager}, especially for sensitive application scenarios such as medicine \cite{karabacak2023embracing}, therapy \cite{cho2023evaluating}, healthcare \cite{cabrera2023ethical, yang2023large}, just to mention a few. 
Given the widespread use of LLMs in such applications, recent studies \cite{karabacak2023embracing, wang2023voyager} have started exploring the challenges and opportunities that emerged from the adoption of LLMs for language generation in these environments. For instance, Cabrera et al. \cite{cabrera2023ethical} conducted a literature review to discuss the bioethical dilemmas related to the use of chatbots in the field of mental health, namely quality of care, access and exclusion, responsibility and human supervision, and regulations and policies for LLM-equipped use. Their review recommend that LLM-equipped chatbots should be developed for mental health purposes, with tasks complementary to the therapeutic care provided by human professionals, and that their implementation should be properly regulated and should have a strong ethical framework.
Analogously, Cho et al. \cite{cho2023evaluating} have investigated the LLM efficacy -- in terms of empathy, communication skills, adaptability, engagement, and the ability to establish a therapeutic alliance -- in interactive language therapy for high-functioning autistic adolescents. Their results highlight the challenges of developing such therapeutic LLM-equipped systems that cannot achieve the depth of personalization and emotional understanding characteristic of human therapists and the importance of ethical considerations in therapeutic contexts.

However, very few studies \cite{brohan2023can, onorati2023creating} have investigated further how the LLM usage in agents has impacted agent language and human perceptions toward the agent. For example, Onorati et al. \cite{onorati2023creating} proposed a robotic application to generate verbal dialogues considering the user’s interests and preferences using LLM to create the conversation. Their main goal was to make users perceive these dialogues as interesting, to avoid disengagement during the interaction with the robot. Their results show that participants positively engage and use the robotic application. Brohan et al. \cite{brohan2023can} proposed a set of pre-trained skills for a robot, which were used to constrain the model to propose natural language actions that were both feasible and contextually appropriate to provide high-level semantic knowledge about the real-world robotic task (e.g., use a vacuum cleaner). 

This work presents the first evaluation of the appropriateness of the language of an LLM-equipped robotic coach that delivered well-being coaching in a workplace context. 

\subsection{Robotic Coaches For Mental Well-being}

Various studies have examined the use of robotic coaches to promote mental well-being, e.g., \cite{axelsson2023robotic, spitale2022affective, spitale2023robotic, jeong2023deploying, bodala2021teleoperated, matheus2022social, shi2023evaluating}. 
Jibo robots were used in a longitudinal study by Jeong et al. \cite{jeong2020robotic} to deliver positive psychology interventions to students in home settings over the course of seven days. It was found that participants gradually grew fond of the robot and experienced improved well-being, improved mood, and readiness to change. 
Bodala et al. \cite{bodala2021teleoperated} evaluated participants' perceptions towards a human coach in comparison to a teleoperated robotic coach across a five-week period. It was concluded that both had favourable feedback, however, the human coach was preferred in terms of likability, intelligence and level of animacy. Additionally, the personality traits of the participants such as \textit{neuroticism} and \textit{conscientiousness} were found to influence how they interacted with the robot. 
Shi et al. \cite{shi2023evaluating} investigated the effect of physical embodiment and personalisation on the user-perceived quality of text-to-speech (TTS) voices for 
mindfulness. Their results showed that the user-personalized TTS voices were able to perform nearly as well as human voices, indicating that user personalisation could be a powerful approach to raise user perception of TTS voice quality.
When Abbasi et al. \cite{abbasi2022can} looked into a robot's potential to evaluate mental well-being problems in children, they found that it was more accurate at identifying likely anomalies than tests that relied on self- and parent-reports. 
Recent studies have looked into the application of robotic coaches in the workplace and public spaces. Spitale et al. \cite{spitale2023robotic, spitale2023longitudinal} conducted a study involving employees of a tech company to interact with two different forms of robotic coaches that delivered positive psychology exercises over 4 weeks. Their results showed that the robot form may impact the perception of the coachees towards the robotic coach. Axelsson et al. \cite{axelsson2023robotic} examined the use of a robotic mindfulness coach at a public cafe over 4 weeks, finding that participants thought the robot was useful as a guiding voice and a focal point, but wanted the robot to be more responsive.
However, these studies have not examined the appropriateness of robotic coach language.

\section{METHODOLOGY}
\label{sec:method}
The ultimate goal of this work is to understand the appropriateness of the language of a robotic coach for promoting mental well-being in the workplace. This is done by undertaking an empirical study and then conducting individual interviews and a focus group with employees who have interacted with the robotic coach. 
This section describes the methodology of the study and the qualitative approach by reporting the participant demographics, the empirical study, the interview and focus group protocols, and the scenarios investigated.

\subsection{Participants}
We conducted an empirical study and individual semi-structured interviews with 17 employees  (7 females, and 10 males, 4 of whom were 18-25 years old, 6 were 26-35 years old, 4 were 36-45 years old, and 3 were 46-55 years old). 
All participants were employees of the Cambridge Consultants Inc. company, where they had taken part in the empirical study (see Sec. \ref{sec:previous_study}). We then asked them whether they were willing to participate in a focus group without specifying to them its ultimate goal.
Eleven participants out of seventeen agreed to attend the focus group: 6 females, and 5 males, 1 of whom were 18-25 years old, 4 were 26-35 years old, 3 were 36-45 years old, and 3 were 46-55 years old.
All participants provided informed consent for their participation and agreed on the usage of their data for scientific research. The focus group design, the protocol, and the consent forms were approved by the Departmental Ethics Committee of the University of Cambridge's Department of Computer Science \& Technology.

\subsection{Study Protocol}
\label{sec:previous_study}

Employees involved in this qualitative research took part in an empirical study with an autonomous and adaptive robotic coach over four weeks at their workplace, as described in detail in \cite{spitale2024NEW}. 

\subsubsection{Empirical Study}
The robotic coach delivered four different positive psychology exercises (one per week), namely savouring, gratitude, accomplishment and one door closes and one door opens. The robot used a LLM-equipped framework for managing the dialogue flow (i.e., we embedded in the robot OpenAI API for natural language processing).
The interaction between each employee and the robotic coach lasted for about 10 minutes and included the following steps:
(1) The robot introduced itself and described the positive psychology practice (just in the first session) and the exercise of the week (e.g., savouring exercise). 
(2) The robot asked the coachee to think about a positive memory from the last week and to share it with them.
(3) The robot listened to the coachee's response.
(4) The robot made a decision on the next step (summarise, ask for a follow-up question, or start a new episode) based on an on-the-fly pre-trained reinforcement learning model described in detail in \cite{spitale2024NEW}.
(5) The robot generated the action according to the decision made in (4) and listened to the coachees' response.
(6) The robot repeated steps (4) and (5) for 8 turns. We decided to fix the number of turns to 8 to ensure that the coaching practice does not last more than 10 minutes.
(7) The robot concluded the session by thanking the employee and reminding them  of the following week’s session.
This procedure was repeated for each session over the four weeks of the study.


\subsubsection{Interviews}
\label{sec:int}
At the end of the last week, we conducted individual semi-structured interviews with all employees (17 in total) in a meeting room of the Cambridge Consultants Inc. headquarters. We asked employees several questions, such as ``What was your overall impression of the robot?'', ``Have you felt understood and listened to?'', ``Would you recommend this robot to a colleague or a friend?'' etc. We concluded the interview by debriefing the coachees, i.e., explaining the main goal of the empirical study, and answering their questions. 

\subsubsection{Focus Group}
\label{sec:fg}

\begin{figure*}[htb!]
    \centering
    \includegraphics[width=0.8\textwidth]{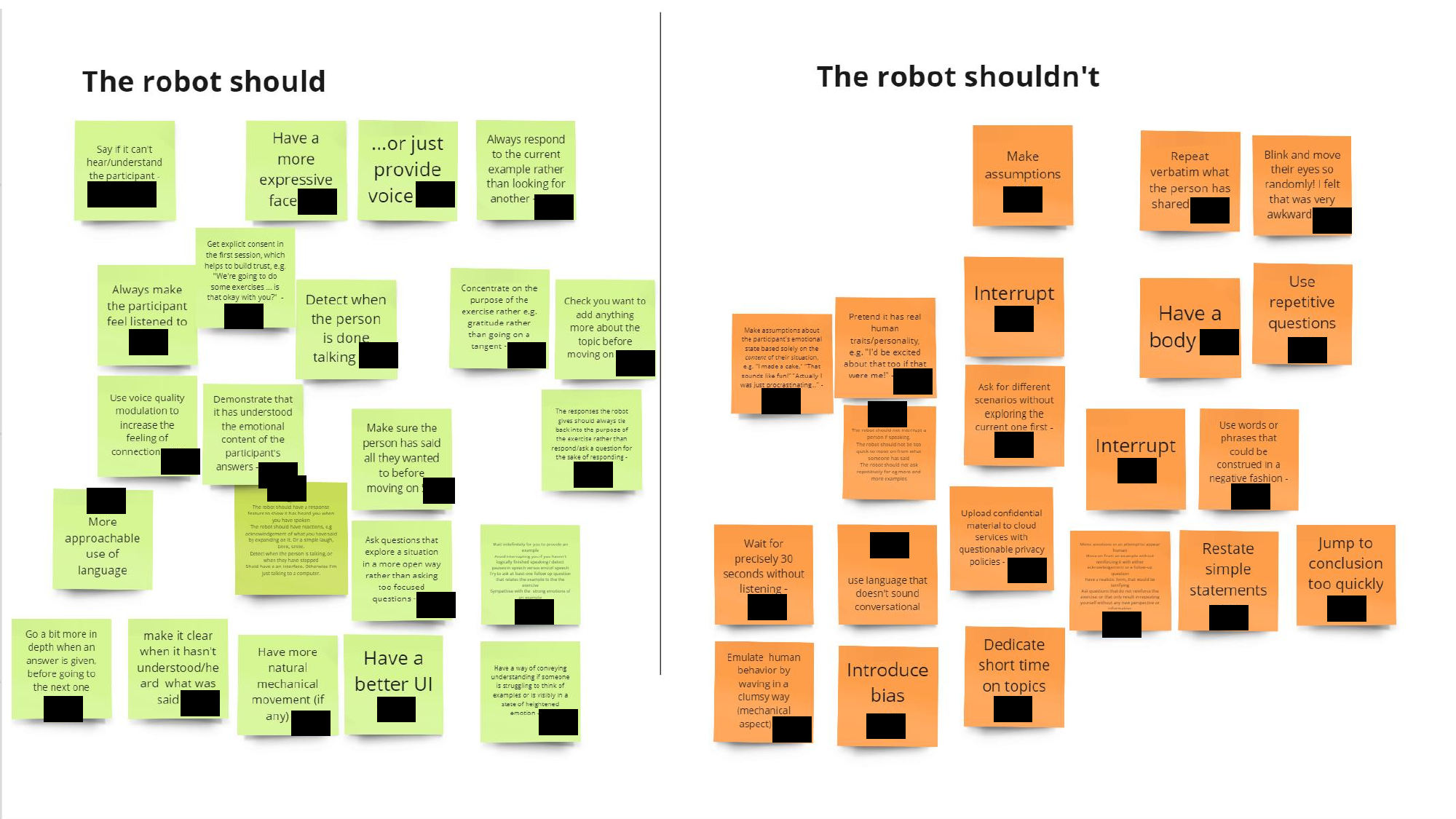}
    \caption{List of should\emph{s} (green post-it on the left-side) and shouldn't\emph{s} (orange post-it on the right-side) identified by the employees regarding the robotic coaches' appropriate behaviours. Participants' initials on the post-its are anonymized using black boxes.}
    \label{fig:guidelines}
\end{figure*}

After two weeks, we conducted a 1.5-hour focus group with the employees online via MS Teams. To facilitate the discussion, we asked participants to use the online tool Miro\footnote{https://miro.com/}. 
Two researchers were present during the focus group: one of them was taking notes, and the other was leading the discussion. 
The focus group included three activities, which were conducted individually on the Miro Board. 
First (\textit{ice-breaking activity}, 10 mins), the researcher asked the employees to post on the Miro board five adjectives to describe the robotic coach. This activity was chosen to better understand the employees' perceptions of the robotic coach as was previously done in \cite{spitale2022socially}. 
Second, the researcher explained to the employees that the main goal of the focus group was to better understand what robotic coaches should say to be perceived as more appropriate to the context and the situation. Then, they described the activity of \textit{scenario evaluations} (40 mins). The employees were presented with seven scenarios on the Miro board, and they were asked to write for each scenario: how they felt about the scenario (``How do you feel?''), the appropriateness score of the robotic coach language in that scenario (from 1-10), how would they modify the robot language ("Behavior modifications"), and why they would modify what the robotic coach says (``Why''). 
Finally, the researcher asked the participants to define a list of \textit{``should''\emph{s} and ``shouldn't''\emph{s}} (20 mins) of the robotic coach language, as depicted in Figure \ref{fig:guidelines}. 
At the end of each of the three activities, the researchers allocated around 10 mins for group discussions with all the participating employees.

\subsection{Scenarios}


In our empirical study, employees shared with the robotic coach various positive episodes from their lives \cite{spitale2024NEW} (e.g., playing instruments, practicing sports) that informed the definition of the workshop's scenarios alongside their grounding in Human-Computer Interaction (HCI) literature. While watching the recordings of the study, we observed situations (i.e., what the coachee shared and how the robotic coach followed up) that we used as scenarios in our focus group, but modified for privacy reasons. 
As a result, we defined seven scenarios that we encountered recurrently in our empirical study by framing them from a Human-Computer Interaction (HCI) perspective \cite{resende2017three}. 
HCI literature \cite{resende2017three} suggests that the research field has had three waves: \textbf{(1)} in the first wave, the human is conceptualised as an object that functions following rigid guidelines, and HCI focused on pragmatic solutions and \textbf{objective measures} (e.g., time, task performance); \textbf{(2)} the second wave added \textbf{subjectivity} and took into account the person in the interaction (besides the functionality of the technology itself) and HCI focused on cognitive science and psychology aspects (e.g., emotion, empathy) during the design of technologies for humans; \textbf{(3)} in the third wave, researchers considered how the technology may reach our everyday lives (homes, privacy etc.), by taking into account \textbf{social-cultural context} (e.g., gender, culture). We applied these waves as the three drivers for designing the seven scenarios as follows:
\begin{enumerate}
    \item \textbf{Wave 1 (Objective Measures) - Efficiency and time pressure}:
        \begin{itemize}
            \item \textit{Scenario 1}: You have just shared that you have been working out recently, and you felt grateful for it. As follow up questions, the robot asks you the questions ``How many times a week are you working out?'', ``When was the last time you worked out?''.

            \item \textit{Scenario 2}: You have just shared one example you were grateful for during the last week. After sharing that, the robot asked you to think about another example you have been grateful for during the last week, giving you 30 seconds to think about it.
            
        \end{itemize}
    \item \textbf{Wave 2 (Subjectivity) - Empathy and emotions:}
    \begin{itemize}
        \item \textit{Scenario 3}: You have just shared that you have become an aunt and you have a new nephew that you are very grateful for. The robot asked you ``What makes you grateful for the birth of your nephew?''.
        \item \textit{Scenario 4:} You have just shared that you have baked a cake as one of your week’s accomplishments. The robot asked you the following question: ``What are the main ingredients for the cake you baked?''
        \item \textit{Scenario 5:} You have just shared that you have been by the beach last weekend and you savoured the moment in which your feet touched the sand. The robot said ``So you just mentioned that you savoured the moment in which your feet touched the sand, what were the senses activated in that moment? The smell, the touch, the sound of the waves?''
    \end{itemize}
    \item \textbf{Wave 3 (Social-cultural Context) - Bias and stereotyping:}
    \begin{itemize}
        
        \item \textit{Scenario 6:} You have just shared that you were grateful that your friend texted you yesterday morning saying that she woke up very energised and she left home after a couple of months of staying on the couch because of her depression. The robot responds: ``I’m really sorry that your friend was feeling depressed. I’m glad to hear that now she is feeling better and that you were able to support her.  Regardless of all your effort, something in her mind should change and you may feel very powerless. What qualities in you made you able to support your friend?''
        \item \textit{Scenario 7:} You have just shared a great accomplishment of a friend of yours at work who successfully delivered a very big project and you are really proud of your friend. The robot assumes that your friend is male and asks you follow-up questions referring to your friend as ``he/him''.
    \end{itemize}
\end{enumerate}

\section{FINDINGS}

This section reports the qualitative evaluation of the findings from the original study and the findings obtained from the interviews and workshop activities approached via a HCI lens. 

\subsection{Interviews}

Following the framework method for qualitative analysis \cite{parkinson2016framework} (i.e., using the three HCI-inspired waves as an analysis framework), we summarise the interview results as follows.

\subsubsection{Wave 1: Efficiency and time pressure} 
Coachees noted that the robotic coach was focusing on the efficiency of their actions rather than going deeper and exploring their feelings, and they also felt pressured by the time constrains driven by the robotic coaches. 
P04 felt that the interaction was like a \textit{"job interview"} during the one door closes and one opens exercise. She shared with the robotic coach a door that closed in her work experience, and the robot kept asking her questions about her work rather than focusing on positive things (e.g., it asked her about time management). P04 also shared in her first week that she has practiced dance. The robotic coach followed up with very practical question, e.g., technical aspects of her dance practice, rather than asking her about how the dance made her feel. 
Analogously, P03 was impressed with how the robot asked a relevant follow-up question about the work task he had described to the robot. However, he expressed disappointment in that the robotic coach asked him specific details about his work task, rather than how his work accomplishment made him feel.
P17 shared with the robotic coach that he played some guitar, and the robot asked several follow-up questions, such as, \textit{"How long have you been practicing?"}, \textit{"How many times a week do you practice?"}, and implying through these statements that the person should have practiced more. The participant found those questions putting ``pressure'' on him, because he ``hasn't actually practiced that much''.
Regarding time pressure, P16 shared that the gratitude exercise \textit{``put him on the spot to think of examples''} of gratitude, and analogously, P11 wished to have more time or at least not feel the pressure of a time out (i.e., the robotic coach was giving the employees 30 seconds to think about each experience before asking them to share). 

\subsubsection{Wave 2: Empathy and emotions} 

Employees also noted that the robotic coach was lacking empathic responses and emotional understanding. Specifically, P15 highlighted that the robotic coach made him \textit{``think of something I would not otherwise''}, but without \textit{``really making him have an emotional response to it''}. P14 explicitly reported that the robotic coach should be \textit{``more empathic''} and also P07 thought that it was \textit{``lacking compassion''} and he didn't feel \textit{``listened to''}.
Again, P06 found the robotic coach not really going deeper in the conversation, but she thought that this could be because the type of example she brought up were \textit{``not very challenging''}. 

\subsubsection{Wave 3: Bias and stereotyping} 

Finally, employees brought up concerns about bias and stereotyping in the interviews. 
P10 highlighted that the robotic coach \textit{``asked really good follow-up questions''}, e.g., by validating her about the difficulty of dealing with toddlers. 
She also mentioned that the robotic coach misheard her when she said that she was giving a presentation and thought ``Dave'' gave the presentation. She found this mistake to exemplify \textit{``gender inequality''}, since she was trying to share with the robotic coach an accomplishment that was hers.


\subsection{Workshop: Ice-breaking Activity}
\begin{figure}
    \centering
    \includegraphics[width = \columnwidth]{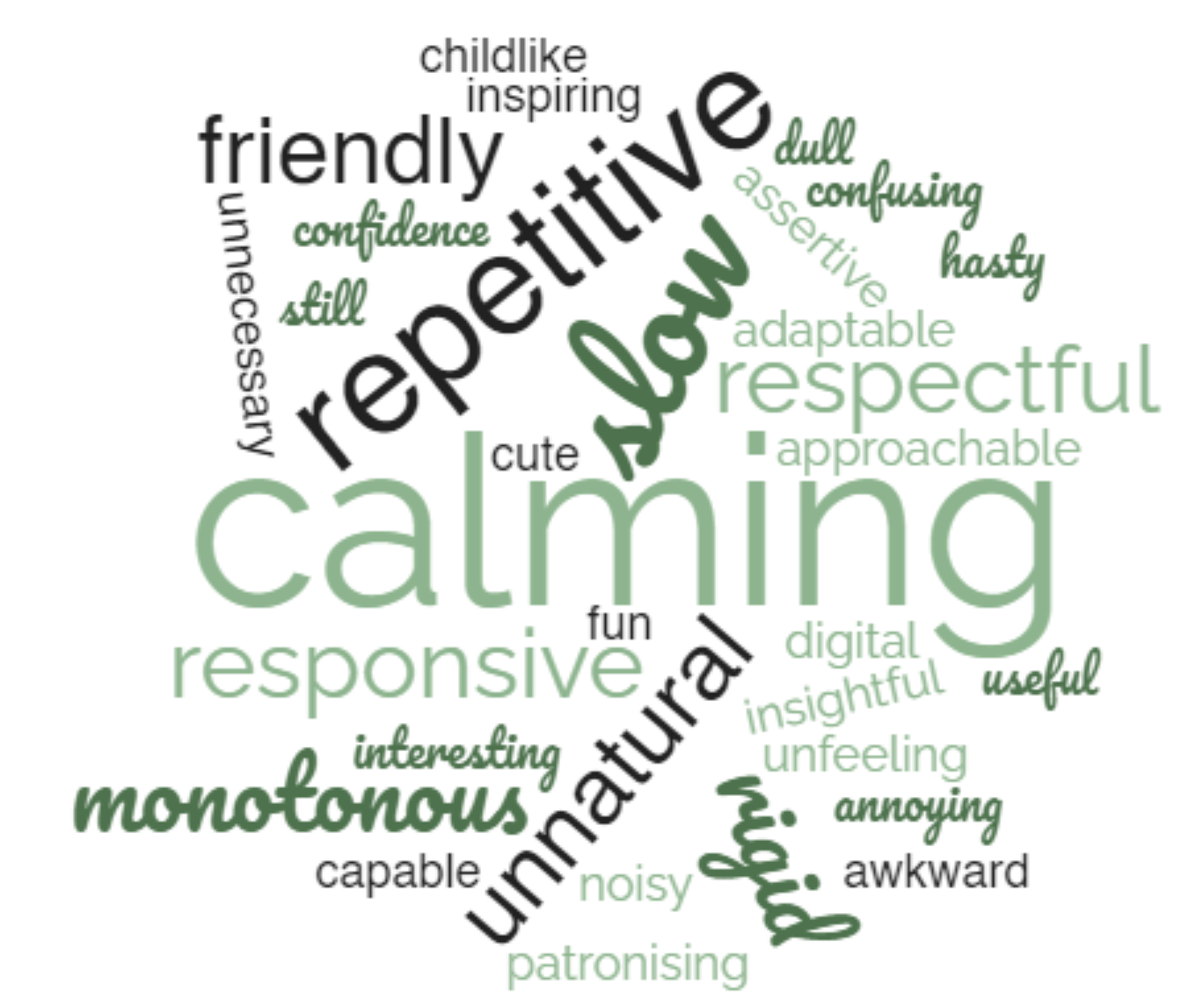}
    \caption{Word cloud of the adjectives listed by the employees to describe the robotic coach.}
    \label{fig:wordcloud}
\end{figure}

Figure \ref{fig:wordcloud} shows the word cloud of the adjectives listed by the employees to describe the robotic coach they have interacted with. Five of the employees used the word \textit{"calming"} to describe the robotic coach because they found the positive psychology exercises delivered by the robotic coach as a beneficial practice to relax and stay calm. P01, P03 and P10 also pinpointed positive aspects of the robotic coach's personality by describing it as \textit{"friendly"}, \textit{"respectful"} and \textit{"insightful"}, and P01 also described its capabilities positively using words like \textit{"responsive"}, and P05 and P03 thought the robot was \textit{"capable"} and \textit{"adaptable"}. However, P05, P08 and P14 perceived the robot to be \textit{"slow"}, \textit{"monotonous"}, and \textit{"repetitive"}, and P01 and P05 found the robotic interaction to be \textit{"unnatural"} and \textit{"rigid"}. 

These results show that employees have both positive and negative opinions of the robot, by describing it as calming and insightful but questioning its capability in terms of naturalness of the interaction and slow pace of the conversation.

\subsection{Workshop: Scenario Evaluation}

\subsubsection{Wave 1: Efficiency and time pressure}
Scenario 1 was interpreted very differently across employees. P03 and P05 thought that the follow up question (i.e., “how many times a week are you working out?”, “when was the last time you worked out?”) of the robotic coach was appropriate, and P05 felt that the robotic coach's question can be \textit{"a natural follow on in a conversation that could provide useful  information"}. P06 believed that the response was appropriate however she suggested to rephrase the follow-up question, e.g. \textit{"Why do you run?"}, or \textit{"Is it something that  you started recently?"}, or \textit{"How do you feel about it?"}.
On the other hand, P04 and P02 found the follow-up question of the robotic coach very inappropriate. P02 felt \textit{"nagged"} and she wondered whether the robotic coach \textit{"is trying to make [her] feel guilty for not doing more [exercise]"}. Also, P04 noted that the robotic coach was not focusing on the gratitude aspect of the exercise but only on facts. She suggested other follow-up questions such as \textit{"What made you feel inspired to exercise?"} or \textit{"Is this something you can plan into your  routine?"}, to make the coachee reflect more why they are grateful for the running experience.
P07 believed that the robotic coach's question could have been appropriate depending on the familiarity between the robotic coach and the coachee, and \textit{"whether the robot has already built rapport and established a relationship"}. He felt that for a first interaction, the presented follow-up question may be perceived as \textit{challenging and confrontational}, while it may be more appropriate later in time.

All employees agreed on the inappropriateness of Scenario 2. They felt \textit{"annoyed"}, \textit{"not listened to"}, \textit{"frustrated"} and \textit{"doubtful"}. This was because they felt the robotic coach was dismissing their first answer and they felt pressured about the timing. P07 thought that it is \textit{"artificial to wait for the full 30 seconds if the participant is ready to speak"}, and also P03 reported that he felt it was stressful to be asked to come up with a new example in only 30 seconds: \textit{"it can make it harder to come up with an example due to the [time] pressure"}. P06 suggested that the robotic coach can ask more follow-up questions regarding the first example, and add asking about another experience 
 only as \textit{"optional"}. P03 would like the robotic coach to \textit{"go deep/follow up in the  discussion for each example, rather than ask for many [examples]"}.

\subsubsection{Wave 2: Empathy and emotions}
Seven employees found Scenario 3 appropriate in terms of the robotic coach's behaviour because they felt \textit{"listened to"} and \textit{"engaged"}. P01 reported that it is very \textit{"nice to elaborate on and reinforce your positive feelings about [the] example"}. P04 suggested that the robotic coach could have also added something like \textit{"family is often a cause of gratefulness"} to show an understanding of the context. Few of them believed that the robotic coach's behaviour was inappropriate. For example, P05 doubted that a \textit{"robot will truly to able to understand it"}, so he found inappropriate that a robotic coach asked such personal questions. P13 also thought that the robotic coach could have dived \textit{"into the reason why it might make you grateful"}.

Four employees found the robotic coach's behaviour in Scenario 4 completely inappropriate and other two considered that behaviour barely appropriate because it did not add any emotional or empathic value to the conversation. P14 believed that asking a follow-up question about the cake's ingredients has \textit{"nothing to do with [the coachee]"}, and suggested that talking about the feelings related to the accomplishment of baking a cake would have been more appropriate behaviour. Analogously, P13 reported that the robotic coach's question \textit{"has no value/benefit for the robot or the  participant"} and suggested that the robotic coach could have asked questions more related to the emotions felt during baking. P10 felt that the robotic coach was \textit{"dismissive"} and found its response to be a \textit{"robotic question"}. P07 felt \textit{"validated"} and \textit{"listened to"} but he found the question \textit{"pointless"} because \textit{"talking about ingredients is  unlikely to elicit a response that leads to rapport/empathy"}. 
In contrast, P05, P04, and P02 found the robotic coach's behaviour appropriate because they felt that it was \textit{"interested in"} what the employee baked. 

Seven employees found the robotic coach's behaviour in Scenario 5 very appropriate. They felt \textit{"listened to"} and \textit{"encouraged to think further about how [the coachee] felt in that moment"}. P04 found the robotic coach's response \textit{"almost perfect"} because the robotic coach was asking to \textit{"return to the moment and think about being there in more detail"}. P02 also suggested that the robotic coach could have asked to \textit{"explore how the sensations made [the coachee] feel"}.
In contrast, three employees found the behaviour inappropriate. For example, P14 felt rushed and he would have preferred that the robotic coach gave him the \textit{"time to respond rather than give [the coachee] options"}.

\subsubsection{Wave 3: Bias and ethics}
All the employees found the robotic coach's behaviour in Scenario 6 very inappropriate. They felt \textit{"confused"}, \textit{"awkward"}, and \textit{"frustrated"}. P03 noted that the robotic coach was inappropriately making assumptions, and analogously P07 noted that the robotic coach was \textit{"making an assumption about [the coachee's] emotional state when [the coachee] has simply described the facts of the situation"}. P02 highlighted that the robotic coach was a \textit{"bit pessimistic"} and assumed that \textit{"any progress the friend has made may be lost"}, and she suggested that instead the robotic coach should have considered the qualities of the coachee and how they have been useful for the friend. Analogously, P13 felt that the robotic coach's answer was \textit{"not constructive"} and she suggested that the robotic coach should ask about what the coachee has done to help their friend and give them support. She also highlighted that this is a very sensitive subject and the robotic coach should be careful when handling such situations that may be \textit{"close to"} many families.

All the employees agreed that in Scenario 7 the robotic coach's behaviour was completely inappropriate because it assumed that the successful friend was a male. They felt \textit{"frustrated"} and \textit{"annoyed"}. P04 was \textit{"pissed off"} by the robotic coach's behaviour because she believed the robot was trained on models that inherited social biases. She suggested that the robotic coach can be trained to use gender neutral pronouns when the gender of the person is not provided. Analogously P10 said that the robotic coach with such behaviour was  \textit{"perpetuating societal stereotypes"}, and he suggested to keep gender neutrality, and make no assumptions in this regard. To avoid making assumptions, P03 suggested to ask more questions rather than fall into stereotypical behaviour hypotheses.

\subsection{Workshop: List of should\emph{s} and should not\emph{s}}

We asked the employees to identify how the robotic coach "should" and "should not" behave while delivering positive psychology coaching sessions. Our findings show that being able to make the participant \textit{"feel listened to"} is one of the main features employees believed the robotic coach should be equipped with. P02 and P07 highlighted that the robotic coach should show cues that suggest that it had heard the coachee when they were speaking. Employees thought that the robotic coach should finish discussing their shared experience by going \textit{"a bit more in depth when an answer is given"}, and by double-checking with the coachee if they have anything to add about the topic before moving on to asking about another experience.
Emotional understanding and expression were other behaviours that employees identified as appropriate for a robotic coach. P07 believed that the robotic coach needs to \textit{"demonstrate that it should understand the emotional content of the participant's answers"} and P03 suggested that it should have a \textit{"more expressive face"}.

Employees highlighted that the robotic coach should not \textit{"make assumptions"} and \textit{"jump to conclusions too quickly"}, to avoid introducing any type of biases. 
In addition, the robotic coach should not repeat \textit{"verbatim"} what the coachee has said to avoid appearing mechnical, and should not interrupt them as that may disrupt the coaching session. 

\section{DISCUSSION \& CONCLUSION}
We collated the qualitative findings from interviews and the focus group, and discuss these results as follows. 

 %
Reflecting the research trend of the \textbf{first HCI wave}, when technologies were measured mostly in terms of objective metrics and performances \cite{resende2017three}, coachees perceived that the robotic coach was too pragmatic and superficial and focused on measurable accomplishments rather than on their feelings. 
Specifically, our findings show that LLM-equipped robotic coach followed up with practical questions and facts without focusing on the positive psychology aspects (e.g., asking why they were grateful for what they shared).
Similar results were found in \cite{vowels2023chatbots} in which health professional interacted with ChatGPT and identified limitations during mental health counselling such as rushing the client, not addressing or evaluating safety concerns, putting clients at risk. 
Again, \cite{limna2023use} highlighted the risks of using LLMs in high education systems in promoting superficial learning rather than deeper explanations. Future work should investigate systematically how to avoid inappropriate responses especially in delicate contexts such as mental well-being coaching.

Coachees also considered the LLM-equipped robotic coach not empathic and not able to emotionally understand their experiences by highlighting the need of putting the subjectivity and psychological aspects at the center of the interaction as for the \textbf{second HCI wave}. Specifically, coachees felt engaged and listened to in accordance to previous studies \cite{vowels2023chatbots}, but not understood emotionally. This result is line with a previous work \cite{belkhir2023beyond} that developed a new way of prompt engineering ChatGPT to enhance the empathic and emotional understanding capabilities of the chatbot. 


In accordance with the \textbf{third HCI wave} that focused on the socio-cultural importance, coachees pinpointed that the LLM-equipped robotic coach should not make assumptions and make statements that reinforce social stereotyping. Specifically, our results show that they perceived the robotic coach was perpetuating social stereotypes and biases such as gender stereotypes.
These findings are in line with a previous study in which undergraduate students interacted with ChatGPT in Arabic language, and ChatGPT generated data
in the context of counseling and mental health was not suitable for Arabic society, customs, traditions, and culture \cite{ajlouni2023students}. This could have been attributed to the bias in the training data as reported previously in a study that focused on bias of ChatGPT in American society \cite{cao2023assessing}.

In summary, our findings suggest that a language-appropriate robotic coach should:
\begin{enumerate}
\item ask deep questions which explore feelings of the coachees, 
\item express and show emotional and empathic understanding of the context, 
\item not make any assumptions without clarifying with follow-up questions to avoid bias and behaviours that enforce stereotypes.
\end{enumerate}
We hope that these results can inform the design of language-appropriate robotic coaches to promote mental well-being in various real-world contexts. 

\section*{ACKNOWLEDGMENT}
\footnotesize
We thank Cambridge Consultants Inc. and their employees for participating in this study. 
\textbf{Funding:} M. Spitale and H. Gunes have been supported by the EPSRC/UKRI under grant ref. EP/R030782/1 (ARoEQ) and EP/R511675/1. 
M. Spitale is partially supported by PNRR-PE-AI FAIR project funded by the NextGeneration EU program.
M. Axelsson is funded by the Osk. Huttunen foundation and the EPSRC under grant EP/T517847/1. 
\textbf{Open Access:} For open access purposes, the authors have applied a Creative Commons Attribution (CC BY) licence to any Author Accepted Manuscript version arising.
\textbf{Data access:} Raw data related to this publication cannot be openly released due to anonymity and privacy issues.




\normalsize
\bibliographystyle{IEEEtran}
\bibliography{ref}

\end{document}